\documentclass[longbibliography,twocolumn,groupedaddress,prX]{revtex4-1}

\usepackage{graphicx}
\graphicspath{{./}}
\begin{document}
\title{Sub-diffusion and population dynamics of water confined in soft environments}
\author{Samuel Hanot}
\email[]{hanot@ill.fr}
\affiliation{Institut Laue-Langevin - 71 Avenue des Martyrs - CS 20156 - 38042 GRENOBLE CEDEX 9}
\author{Sandrine Lyonnard}
\email[]{sandrine.lyonnard@cea.fr}
\author{Stefano Mossa}
\email[]{stefano.mossa@cea.fr}
\affiliation{Univ. Grenoble Alpes, INAC-SPRAM, F-38000 Grenoble, France}
\affiliation{CNRS, INAC-SPRAM, F-38000 Grenoble, France}
\affiliation{CEA, INAC-SPRAM, F-38000 Grenoble, France}
\date{\today}
\begin{abstract}
We have studied by Molecular Dynamics computer simulations the dynamics of water confined in ionic surfactants phases, ranging from well ordered lamellar structures to micelles at low and high water loading, respectively. We have analysed in depth the main dynamical features in terms of mean squared displacements and intermediate scattering functions, and found clear evidences of sub-diffusive behaviour. We have identified water molecules lying at the charged interface with the hydrophobic confining matrix as the main responsible for this unusual feature, and provided a comprehensive picture for dynamics based on a very precise analysis of life times at the interface. We conclude by providing, for the first time to our knowledge, a unique framework for rationalising the existence of important dynamical heterogeneities in fluids absorbed in soft confining environments.       
\end{abstract}
\maketitle
\section{Introduction}
\label{sect:intro}
Nanoconfinement and interfacial interactions profoundly alter thermodynamical, structural and transport properties of confined fluids. A molecular-level understanding of these modifications is essential in a diversity of fields, ranging from biology~\cite{schwabe1997role} to nanoscience~\cite{hummer2001water}. A situation of paramount interest is that of water adsorbed in natural or synthesized soft media, such as biomolecules, macromolecular gels, polymers, self-assembled and active materials. In particular, it is believed that the function-structure relationship in these systems is primarily correlated to the amount of the adsorbed fluid and to its peculiar properties. Water in the vicinity of soft charged walls is expected to show significant deviations from the bulk behaviour, e.g., distortions of the hydrogen bonds network and anomalous diffusion character, which might in turn profoundly impact the final morphology and functional behaviour of the confining media. A deeper insight of such complex fluid-matrix interplay can be obtained by using numerical tools. Here, we use Molecular Dynamics simulations of a self-assembling model system to demonstrate and rationalize the general existence of spatially heterogeneous dynamics in a fluid adsorbed in soft confining environments.

The generic situation of soft mobile confining media has been less explored than that of hard confining materials. Detailed studies on confinement in highly stylised immobile matrices have been described, which aim at understanding the universal phenomena originating from both the detailed nature of the fluid/matrix interactions and the restricted character of the available space. These include, together with many other systems, simulations of water in carbon nanotubes~\cite{Liu2005}, confined between two smooth hydrophobic plates~\cite{PhysRevE.72.051503} or patterned walls~\cite{PhysRevLett.102.050603}, or in the case of patterned hydrophilic and hydrophobic walls~\cite{doi:10.1021/jp065419b}. In contrast, most studies on water confined in (disordered) soft matrices are very specific, and focus on particular confining macromolecules, without any concern for generality. These investigations include, among others, neutron scattering and molecular dynamics simulations of tau and MPB proteins~\cite{Schiro2015}, dynamics of water in reverse surfactant micelles ~\cite{harpham2004water}, or simulation of polymer electrolyte membranes~\cite{Devanathan2007a, Devanathan2007}. These molecules, be they of biological or synthetic origin, are very complex, and one finds extremely difficult to separate properties that are specific to the particular chemistry of the confining matrix, from those that are general and emerge immediately when water is in contact with {\em any} soft confining media. Still, in general we expect that the main features of water transport in these systems must conform to some universal description, dictated by the principles of statistical mechanics.
\begin{figure*}[t]
\centering
\includegraphics[width=1.0\textwidth]{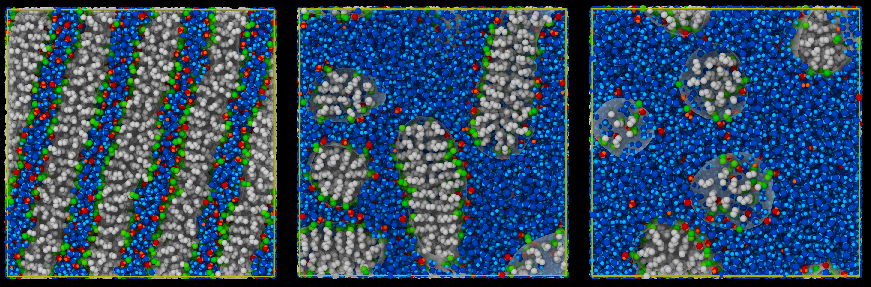}
\caption{
Typical simulation snapshots of the ionic surfactants model studied here, at hydration levels $\lambda=4, 16, 32$, from left to right. Lamellar, cylindrical and micellar self-organised phases are evident. The hydrophobic section of the surfactant macromolecules is represented by the grey beads, the charged hydrophilic units by the green ones. Water molecules are indicated in blue, the hydronium cations in red. These configurations are examples drawn from the larger set of matrix templates we have considered in our investigation of the dynamics of water molecules confined in self-assembled soft environments. 
}
\label{fig:snaps}
\end{figure*}

It is well understood that water transport around proteins is highly anomalous~\cite{Yamamoto2014a,Yamamoto2013,Bizzarri2002,Pizzitutti2007}, with sub-diffusive water molecules playing a crucial role in the function of the protein. Notwithstanding this well-established observation, the existence of sub-diffusion in synthetic materials is scarcely discussed. Surprisingly, most studies on synthetic materials focus on very complex simulations, with very detailed force fields, but they generally do not capitalise on this modelling accuracy to provide detailed and up-to-date analysis of the transport properties of the ab(d)sorbed fluid. For instance, the dynamics of water in surfactants is treated only in the reverse micelle region of the phase diagram~\cite{hahn1996single}, and most studies consider water adsorbed in proton-conducting fuel cells ionomers as a Fickian diffusive system~\cite{Sunda2013,Devanathan2007a}. Yet, the functional properties of these materials are also determined by the mobility of water confined in soft ionic nanochannels. In addition, from the point of view of the elementary water-matrix interactions, ionomers, for instance, substantially resemble biological matter: {\em i)} they both contain hydrophilic and hydrophobic parts, {\em ii)} neutral and charged groups, {\em iii)} water is confined at the nano-scale, and {\em iv)} it evolves in the presence of soft charged interfaces. From the point of view of the function, these materials are also alike. In one case we must understand how particular species reach specific active sites on proteins or lipidic membranes, and how this relates to the surrounding water. In the other, we aim at grasping the mechanisms behind transport of ions within the ionic channels, which depend on the shuttling efficiency of the water molecules. 

Here, we try to elucidate these issues by investigating the transport properties of water molecules confined in self-organised ionic surfactants phases, across a large range of hydration levels. We have selected a perfluorosulfonic surfactant molecule to insure perfectly phase-separated morphologies with well-defined sharp interfaces, arising from the extremely hydrophobic character of the tail and important super-acidity of the head group. We have considered the numerical model for ionic surfactants of Ref.~\cite{Hanot2015}, which is able to stabilise well-ordered self-assembled phases, ranging from lamellae to reverse micelles at low and high water content, respectively. We have systematically followed the constrained evolution of water adsorbed in these mobile soft charged environments, on extended time scales. On the basis of data sets of unprecedented statistical accuracy, we demonstrate that it is possible to rationalise in a comprehensive picture both the {\em average} sub-diffusive character of the dynamics, due to the subtle interplay between confinement at the nano-scale and direct interaction with interfaces with different topological properties, and the {\em spatially-resolved} heterogeneous behaviour in different regions of the ionic channels~\cite{damasceno2013inhomogeneous}. We also believe that our work both highlights the similarities in transport properties of systems of biological interest and synthetic materials, and provides physical insight for a convincing interpretation of phenomenological modelling of experimental data.
\section{The model}
\label{sect:model}
In Ref.~\cite{Hanot2015}, we described a coarse-grained numerical model for (explicit) water confined in ionic surfactants phases. We considered a three-components mixture, formed by the surfactant macro-molecules, the polar solvent (water), and the solvated counter-ions (hydronium). We chose a united-atoms representation for the surfactant molecule, inspired by the model for the side-chain of Nafion of Ref.~\cite{allahyarov2008simulation}. The hydrophobic part of the molecule is represented by a chain of seven Lennard-Jones (LJ) beads of neutral electric charge, each one coarse-graining an entire CF$_2$ group. Similarly, the head group is schematised by two charged LJ beads, one for the sulphur atom S and one for the O$_3$ group, with a total charge $q=-e$. The covalent bonds are modeled by quadratic potentials. The non-bonded interaction forces were accurately parametrised, in order to generate an extremely neat phase separation between the hydrophobic matrix and the ionic channels, at all investigated water contents. The popular SPC/E model was used for the water molecules. All details can be found in~\cite{Hanot2015}.

\begin{figure}[t]
\centering
\includegraphics[width=0.5\textwidth]{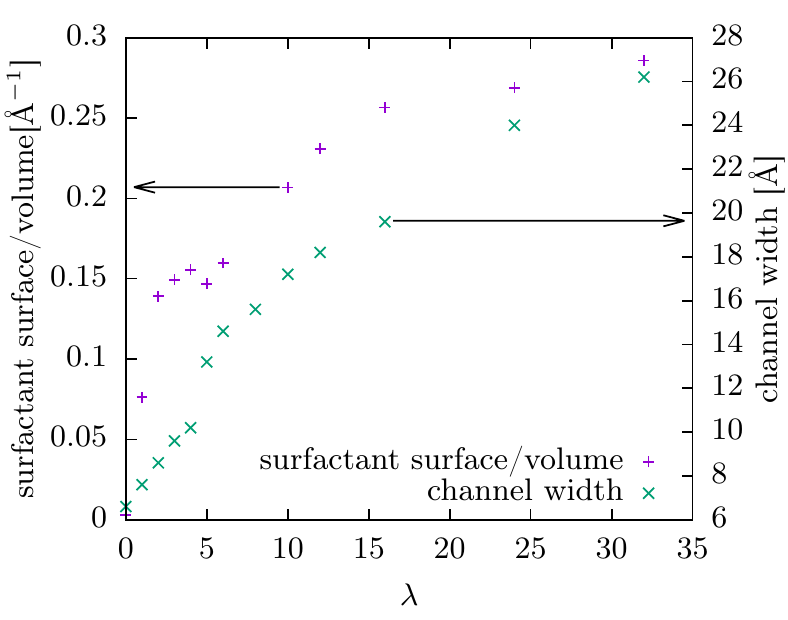}
\caption{
Variation with the water content, $\lambda$, of the average size of the self-assembled ionic channels and the extension of the generated hydrophobic/hydrophilic interfaces, normalised  
to the volume occupied by the surfactant macromolecules. Data for the confining size are similar to those reported in~\cite{Hanot2015}, re-evaluated from the new system snapshots. The arrows indicate the axis relevant for each data set.
}
\label{fig:matrix}
\end{figure}

The water content $\lambda$, i.e., the number of water molecules per ionic surfactant macromolecule, influences the macroscopic self-assembly and generates a sequence of phase changes, encompassing lamellar, cylindrical and micellar phases by increasing hydration. In Fig.~\ref{fig:snaps} we show typical snapshots of the simulated systems, at the indicated values of $\lambda$, increasing from left to right. Also, in Fig.~\ref{fig:matrix} we show the $\lambda$-variation of the average confining size and the surface of the generated hydrophilic/hydrophobic interfaces. (This last quantity has been calculated by determining a polyhedral mesh around the beads pertaining to the surfactants~\cite{stukowski2014computational}, and is normalised to the total volume it contains.) Both data sets generally increase with $\lambda$ following, however, non-trivial patterns which also highlight the boundaries between the different phases. The behaviour of the confining size can be attributed to the system swelling upon increasing water loading. A more subtle argument, based on an increasing curvature radius of the interfaces, is needed to fully explain the details of the interfaces growth~\cite{Hanot2015}. 

The interest of this model in the present context is two-fold. First, it constitutes a very effective tool to autonomously generate very different confining environments, modulating both  the typical confinement sizes and the extent and topological features of the formed interfaces. Second, the extremely neat phase separation provides us with the possibility of precisely identify and characterise the hydration layer adjacent to the interface with the hydrophobic matrix, complementary to the volume occupied by water of relatively higher bulk-like character, far from the boundaries. Both points are crucial for rationalising the dynamical behaviour of the adsorbed fluid, as it will be clear below. 
\section{Simulation details}
\label{sect:simulation}
We have used the parallel high-performance computing tool for Molecular Dynamics simulations LAMMPS~\cite{plimpton1995fast} to generate the trajectories. In order to assure an optimal  thermalisation of the systems at all hydrations, we reinitialised the simulation boxes starting from the self-assembled configurations analysed in~\cite{Hanot2015}. We considered 9 values of the hydration $\lambda$ ranging in the interval $\left[ 1,32\right]$, which encompass all interesting regions of the phase diagram. A pure water system (bulk) trajectory was also generated, as a reference. Note that the mobility of the surfactant macro-molecules lifts the need for an additional average over multiple realisations of the confining matrix. We verified in a few cases, however, that equivalent results are found when preparing the system in different initial conditions.    

At each $\lambda$, we produced a trajectory spanning a total time $\tau_{\text{max}}=10$~ns, a dynamical window sufficiently extended to allow comfortably observing the very slow dynamics in the most confined cases. The time step used for the numerical integration of the equations of motion is $\delta t=$~1 fs. We worked in the constant temperature and pressure (NPT) ensemble, with $T=300$~K and $P=1$~atm, with the characteristic time of both thermostat and barostat set to $\tau_{\text{T}}=\tau_{\text{P}}=1$~ps. We dumped complete systems configuration with a frequency of $10$~ps, for subsequent analysis.
\section{Results}
\label{sect:results}
In this Section we describe our computer simulation data. We have fully characterised the dynamics of water molecules adsorbed in the surfactants nano-phases at the different hydration levels, in terms of mean-squared displacements and intermediate scattering functions. We have next completed our analysis by probing well-designed correlation functions associated to the typical life-times of water molecules in the hydration layer, in contact with the confining hydrophobic matrix.
\begin{figure}[t]
\centering
\includegraphics[width=0.5\textwidth]{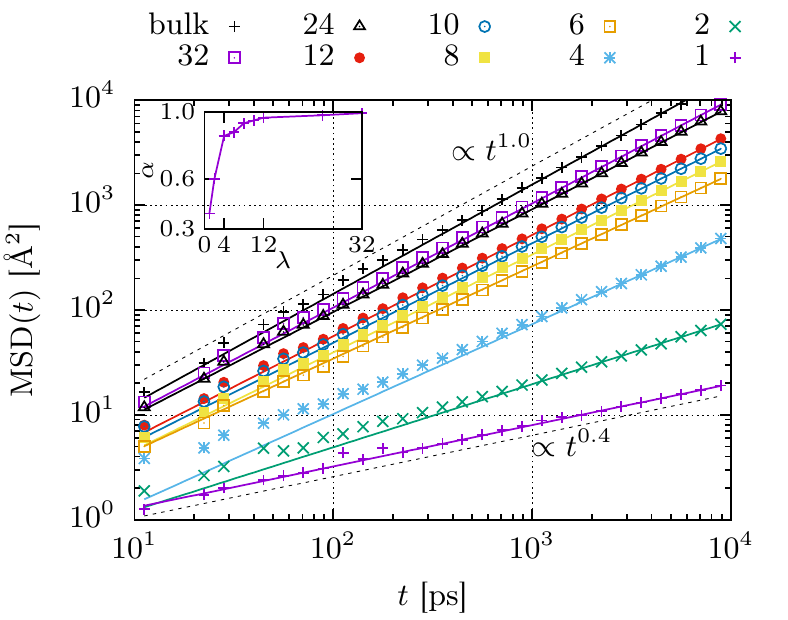}
\caption{
{\em Main panel:} Mean-squared displacements $\langle r^2(t) \rangle$, calculated from the time-dependent coordinates of the water oxygen atoms. The solid lines are the power-law fits described in the main text, the dashed lines indicates the limiting Fickian (top) and highly sub-diffusive (bottom) limits, corresponding to $\alpha=1$ and $0.4$, respectively. {\em Inset:} $\lambda$-dependence of the exponent $\alpha$, determined from $\langle r^2(t) \rangle\propto t^{\alpha}$.
}
\label{fig:subdiffusion}
\end{figure}
\subsection{Evidence of sub-diffusive behaviour}
\label{subsect:subdiff}
By using the produced system trajectories, we first computed the water molecules mean-squared displacement (MSD) 
\begin{equation}
\langle r^2(t) \rangle = \frac{1}{N_w}\langle \sum_{i=1}^{N_w} \left | \mathbf{r}_i(t) - \mathbf{r}_i(0) \right |^2 \rangle,
\label{eq:msd}
\end{equation}
where $N_w$ is the number of water molecules, and $\mathbf{r}_i(t)$ is the position vector of the oxygen atom pertaining to molecule $i$ at time $t$. In Fig.~\ref{fig:subdiffusion} (top) we show our data at the investigated various hydration levels (higher on top), together with the results for bulk water. We observe two interesting features: first, by decreasing the hydration, the mean-squared displacement at fixed time decreases, as generically expected due to an increasing crowding associated to the surfactant units. Second, and more intriguing, while the data-points for the bulk fall on a straight line of unitary slope (the linear dependence expected for water at ambient conditions), this slope progressively decreases with the hydration, indicating a power-law behaviour with an exponent less than one. This is characteristic of {\em sub-diffusion}~\cite{metzler2014anomalous}. 
\begin{figure}[b]
\centering
\includegraphics[width=0.5\textwidth]{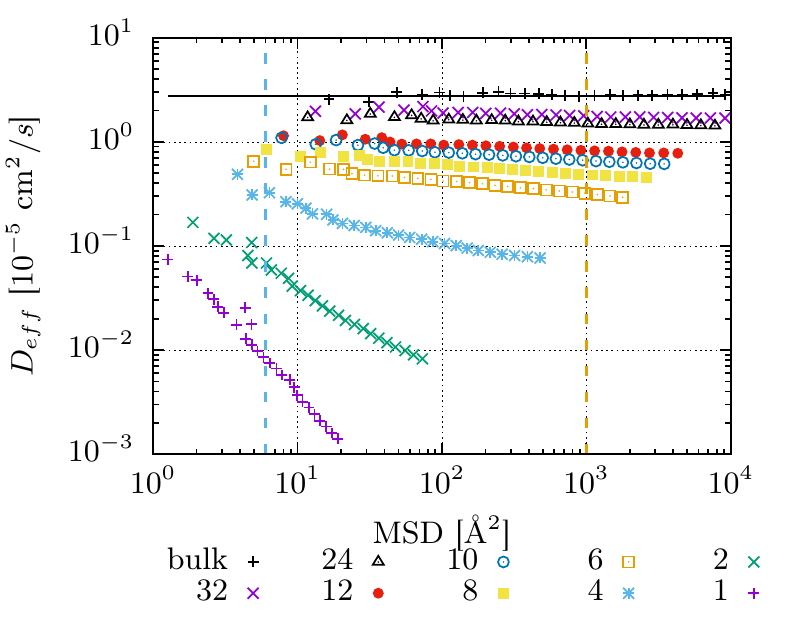}
\includegraphics[width=0.5\textwidth]{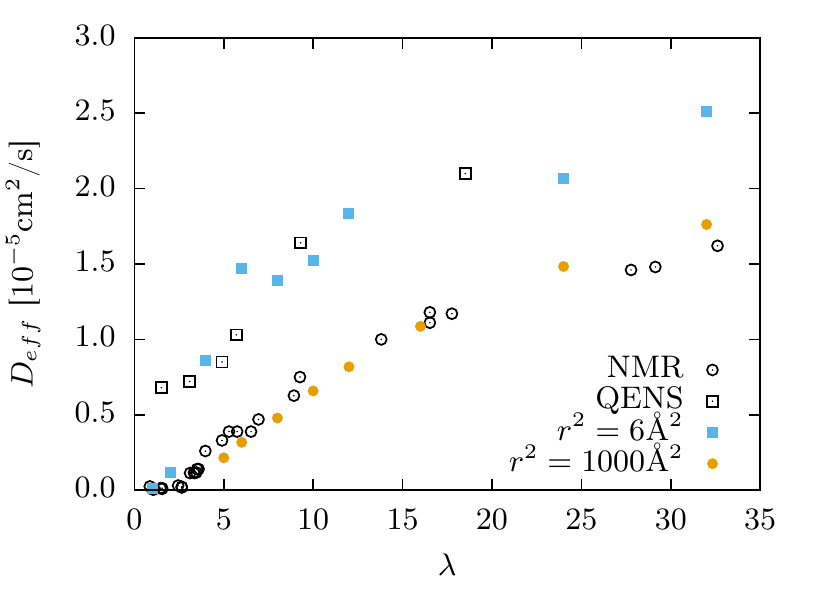}
\caption{
{\em Top:} The effective diffusion coefficient, $D_{\text{eff}}$, computed from Eq.~(\ref{eq:Deff}), plotted parametrically against $\langle r^2(t)\rangle$. The solid black line is the result in the bulk for the considered water model, corresponding to $D\simeq 2.74\times 10^{-5}$ cm$^2$/s. The dashed vertical lines, indicate the (squared) length scales $\simeq 6$ and $\simeq 1000$~\AA$^2$ used to select the values of $D_{\text{eff}}$ shown in the bottom panel, as discussed in the text. {\em Bottom:} The values for $D_{\text{eff}}$ selected by the two verticals lines of the above panel are shown as closed symbols. We also show the diffusion coefficients obtained by PFG-NMR and QENS spectroscopy~\cite{berrod2015qens}, which probe the dynamics on long and short length scales, respectively.
}
\label{fig:diff15A2}
\end{figure}

To quantify this feature, we fit the MSD at long times to a power-law of the form $\langle r^2(t) \rangle\propto t^{\alpha}$, with $0<\alpha=\alpha(\lambda)\le 1$. We show the results of the fit as lines in Fig.~\ref{fig:subdiffusion}, and the values of the exponent $\alpha(\lambda)$ in the inset of the same figure. We observe that $\alpha$ varies slowly from a value close to $1$ at $\lambda=32$, corresponding to the expected Fickian diffusion, to $\alpha\simeq 0.85$ at $\lambda=5$, close to the boundary of the lamellar phase. This behaviour seems to correlate with the corresponding decrease of the available interface shown in Fig.~\ref{fig:matrix}. Next, $\alpha$ abruptly decreases to a value $\simeq 0.4$ at $\lambda=2$, certainly controlled by the extreme confining sizes at almost constant available interface. Note that this value is even lower than $\alpha=0.5$, corresponding to the case of single-file diffusion~\cite{hahn1996single}, where molecules are organised in uni-dimensional configurations, with a vanishing position swap probability for adjacent molecules. Altogether, these data demonstrate that our model allows to continuously tune the degree of sub-diffusion by varying the hydration level, interpolating between the extreme cases of Fickian and single-file-like diffusion.

In this context, the fluctuation-dissipation Einstein relation relating the mean-squared displacement to the diffusion coefficient, does not hold in the usual form. We can, however, define a time and length-scale dependent generalised (effective) diffusion coefficient, $D_{\text{eff}}$, from the local slope of the mean squared displacement as~\cite{allen1989computer}, 
\begin{equation}
\partial_t \langle r^2(t)\rangle = \frac{\alpha}{t}\langle r^2(t)\rangle = 6 D_{\text{eff}}.
\label{eq:Deff}
\end{equation}
The effective diffusion coefficients and the corresponding $\langle r^2(t)\rangle$ are plotted parametrically in Fig.~\ref{fig:diff15A2}~(top). Consistently with the data of Fig.~\ref{fig:subdiffusion}, the most diluted cases exhibit diffusion coefficients that are rather constant and close to the value of the bulk. In contrast, upon decreasing hydration, $D_{eff}$ is strongly suppressed, with this effect becoming increasingly evident as the measured $\langle r^2\rangle$ increases. 

An important remark is in order at this point. The data shown in Fig.~\ref{fig:diff15A2}~(top) provide us with a unique rationalisation of the diffusion properties probed by different experimental spectroscopic techniques. We illustrate this point in Fig.~\ref{fig:diff15A2}~(bottom), where we compare our results with the values obtained by Pulsed-Field Gradient Nuclear Magnetic Resonance (PFG-NMR )~\cite{berrod2015macromolecules} and Quasi-Elastic Neutron Scattering (QENS)  experiments on gels of PFOSA surfactants~\cite{berrod2015qens}. We recall that PFG-NMR (open circles) measures the self-diffusion of water on large distances, of the order of the micrometer ($\mu m$), with a typical time scale of a few milliseconds, while QENS (open squares) probes proton displacements of the order of a few angstroms (\AA), on time scales ranging from 1 to typically a few hundreds picoseconds. 

Following an uncommon protocol, we sample our data at fixed values of $\langle r^2 \rangle=(l^*)^2$, compatible with these length scales (vertical dashed lines in the top panel of Fig.~\ref{fig:diff15A2}). Obviously, the $\mu m$ range explored by PFG-NMR is far beyond the limit of our simulations. We note however, that at least for the highest hydration levels, $D_{\text{eff}}$ is almost length-scale independent for $\langle r^2(t)\rangle\ge 10^3$~\AA$^2$. This allows us to safely fix an observation threshold at $l^*\simeq 32$~\AA. Similarly, we have determined a short length-scale value of $D_{\text{eff}}$ with $l^*\simeq 2.4$~\AA, which is compatible with the typical nano-scale distances probed by QENS. Interestingly, the simulation points nicely reproduce the hydration dependence of both experimental data sets, without resorting to any additional consideration about different dynamical processes on different time scales. Our model therefore seems to provide not only an adequate dynamical behaviour compared to the real system, but also a framework to interpret experimental results, without needing any fitting procedure (beside the determination of $\alpha$). In the following we will demonstrate that it also is an efficient tool to pinpoint the origin of the sub-diffusive behaviour described here.
\begin{figure}[t]
\centering
\includegraphics[width=0.5\textwidth]{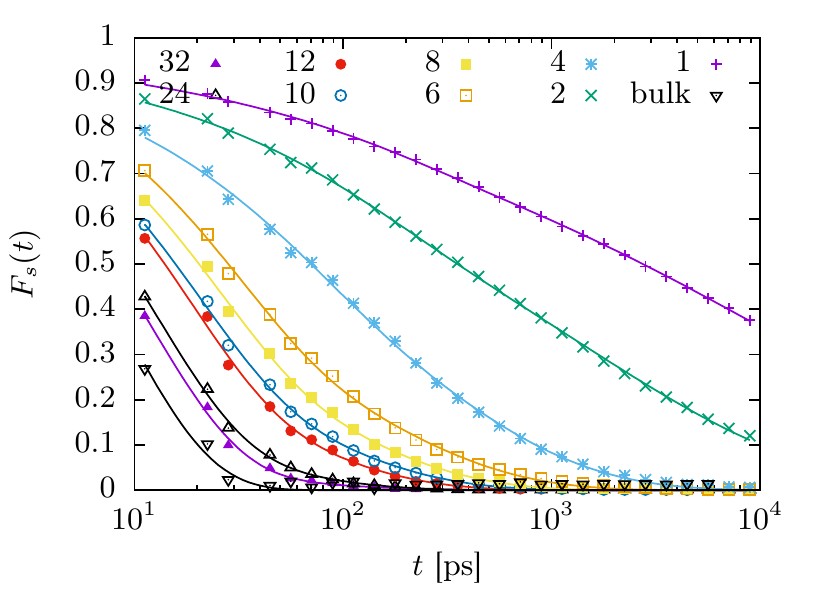}
\caption{
The self-intermediate scattering functions, $F_s(q,t)$, calculated from the oxygen atoms of the water molecules via Eq.~(\ref{eqn:fqt}) (symbols). These data correspond to $q^*=0.7$~\AA$^{-1}$, at the indicated values of $\lambda$. We also show as solid lines the best-fits to the data, according to Eq.~(\ref{eqn:fqt-fit}). These results strongly support the hypothesis of the existence of two dynamical populations, as discussed in details in the main text.}
\label{fig:fs-2exp}
\end{figure}
\subsection{Existence of two dynamical populations}
\label{subsect:populations}
Additional information on the dynamics of the water molecules is carried by the (self-)intermediate scattering function,
\begin{equation}
F_{s}(q,t) = \frac{1}{N_w} \sum_{i=0}^{N_w} \left \langle e^{i\, \mathbf{q} \cdot \left( \mathbf{r}_i(t)-\mathbf{r}_i(0) \right) } \right \rangle_{|\vec{q}|=q},
\label{eqn:fqt}
\end{equation}
where $\langle \rangle_{|\vec{q}|=q}$ is the spherical average over wave vectors of modulus $q$. A strong connection exists in the bulk between the $F_{s}(q,t)$ and the diffusion coefficient. We expect it to partially survive, even if in some modified form, in confinement conditions. The Fourier transform of Eq.~(\ref{eqn:fqt}) is the dynamical structure factor, $S(q,\omega)$, measured in the QENS experiments mentioned above. Here we focus on the value $q^*=0.7$~\AA$^{-1}$~\footnote{At this $q$-value the effect of confinement is clearly observed in the QENS spectra analysis. More precisely, the Half Width at Half Maximum (HWHM) of the Lorentzian-shaped fitting line starts to deviate from a classical random jump diffusion behaviour and enters into a plateau-like regime, attributed to spatial restrictions}. We show our data at all the indicated values of $\lambda $ in Fig.~\ref{fig:fs-2exp}, together with the reference case of the bulk.  We observe that the decay of $F_s(q,t)$ becomes progressively slower as the hydration is decreased, and that the general shape of these functions is quite complex. In the following we quantify these observations. 

The long-time behaviour of $F_s(q,t)$ for bulk molecular liquids is often represented as a stretched exponential of the form $\propto e^{-(t/\tau)^\beta}$. Here $\tau$ is the typical time scale associated to the observed relaxation, and $0<\beta\le 1$ is the stretching parameter. The case $\beta=1$ corresponds to the simple exponential behaviour, $\beta<1$ describes the case where non-trivial processes concur to the average relaxation of the correlation function. As one could expect from the extremely complex behaviour of $\langle r^2(t)\rangle$, this simple functional form alone is not able to accurately reproduce our data, especially at low $\lambda$ where the degree of sub-diffusivity is higher. 

The simplest alternative choice is to assume the existence of two molecular populations characterised by different dynamical properties~\cite{volino2006gaussian,perrin2007quasielastic,berrod2015qens}. In this quite crude representation, a subset of the total number of water molecules evolve diffusively, not very differently than those in bulk conditions, on a relatively fast time scale. The remaining part, in contrast, include all the complex effects coming from confinement in the surfactant phase, evolving on slower time scales. The precise features of the two populations should intuitively depend on the details of the confinement environment and, ultimately, on $\lambda$. Given the additivity of Eq.~(\ref{eqn:fqt}), we can therefore consider the very simple model,
\begin{equation}
F_s(q,t) = f_{\text{SD}}\; e^{-\left(\frac{t}{\tau_{\text{SD}}}\right)^{\beta}} + (1-f_{\text{SD}})\; e^{-\left(\frac{t}{\tau_{\text{D}}}\right )},
\label{eqn:fqt-fit}
\end{equation}
where a simple and a stretched exponentials encode the dynamics of the diffusive (D) and sub-diffusive (SD) water molecules, respectively~\footnote{Note that for molecular liquids the functional form of Eq.~(\ref{eqn:fqt-fit}) is sometimes used for fitting $F_s(q,t)$ in the entire time range, with the simple and stretched exponential well representing the short (caging) and long (structural relaxation) time behaviour, respectively. Here, these two terms are needed for convincingly reproducing the structural relaxation alone, for times $t\ge 10 $~ps.}. Since this latter population is the slowest, we can expect $\tau_{D}<\langle\tau_{SD}\rangle$, where the average time scale associated with the sub-diffusive population is $\langle\tau_{SD}\rangle= \left(\tau_{SD}/\beta\right)\Gamma\left(1/\beta\right)$, with $\Gamma(x)$ the Gamma function. We can also obtain the fraction of diffusive molecules, as $f_D=(1-f_{SD})$.

We were able to obtain excellent fits to our data with this model, that we show as solid lines in Fig.~\ref{fig:fs-2exp}. The obtained values of $f_{SD}$ span the entire possible variation range, assuming a value of $0.05$ at the highest $\lambda$ and of $0.95$ at the lowest available hydration level. This variation is followed by $\beta$-values ranging from $0.6$ to $0.3$. These observations point to an increasing modification of the dynamics compared to the bulk unconstrained case, and to a very probable connection between the sub-diffusive average character of the dynamics and the existence of molecular populations, characterised by heterogeneous dynamical properties. We postpone a much more elaborated discussion of the calculated values for both the time scales and $f_{SD}$ to the following (see Fig.~\ref{fig:fits}).
\subsection{The interface as the source of sub-diffusion}
\label{subsect:interfaces}
The self-assembled confining environments considered here are characterised not only by variable size of the ionic domains available for transport (Fig.~\ref{fig:matrix}), but also by the variable topological features of the formed interfaces, which evolve from the almost perfect planar configurations of the lamellar phase, to configurations of increasing convexity in the cylindrical and micellar phases. It is natural to expect that beside the obvious role played by the extremely tight confining sizes at very low $\lambda$, these well-developed interfaces significantly perturb the dynamics of the adsorbed fluid molecules, participating to the generation of the observed sub-diffusion. More specifically, the surfactant sulfonic acid groups are hydrophilic and negatively charged, resulting in a strong attractive potential between the interface and fluid molecules within the range of these interactions. This leads to the hypothesis that the fluid molecules in the vicinity of the interface can be trapped at the interface, which reduces their possibility of diffusing at the normal rate. This constrain would next be transferred to successive fluid layers, with a strength reduced as the distance from the interface increases. 

On this basis, we now focus on the dynamical properties of the fluid adsorbed at the interface {\em only}, by selecting the sub-set of fluid molecules comprised in the first hydration layer of the surfactant heads. We therefore define the binary presence function, $p(\mathbf{r})$, for a water molecule at position $\mathbf{r}$: $p(\mathbf{r})=1$ if it exists a charged surfactant head at $\mathbf{r}_0$, such that $|\mathbf{r}_0-\mathbf{r}|\le r_c$, and zero otherwise. In what follows we have fixed $r_c=4$~\AA. Applying this operator to the position of the oxygen atoms for every snapshot of our trajectory, we have obtained $N_w$ series of $N_s$ binary values, $\{p_{i,n}\}$. Here, $N_s$ is the number of system snapshots, each corresponding to a simulation time $t_n=n\,\delta t$, with $n=0,\ldots,N_{s}-1$, and $i=1,\ldots, N_w$.
\begin{figure}[t]
\centering
\includegraphics[width=0.5\textwidth]{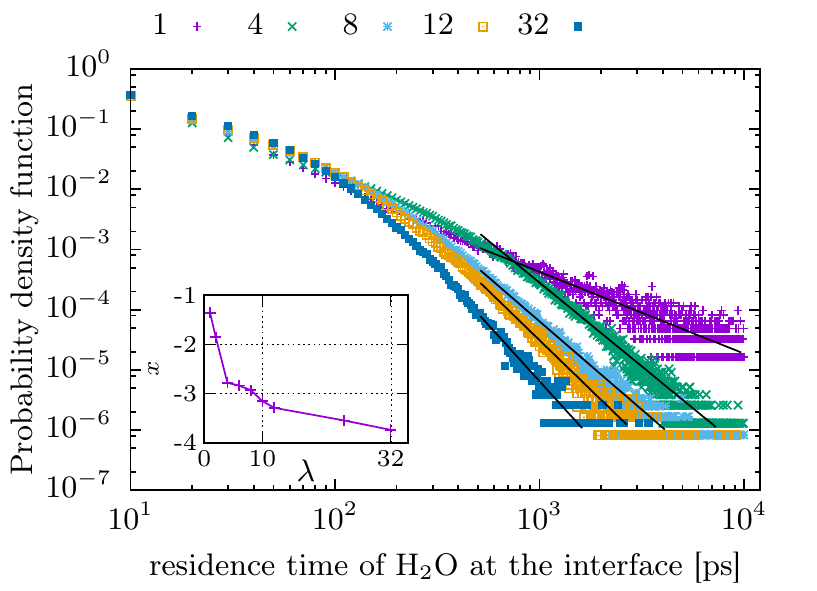}
\includegraphics[width=0.5\textwidth]{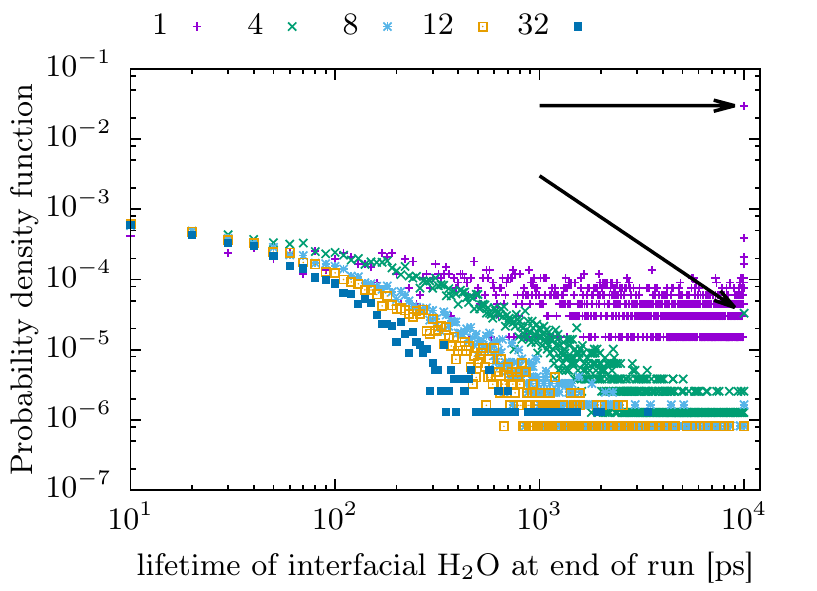}
\caption{{\em Top:} Probability distributions of the residence times observed during the 10ns simulation run. The long-time behaviour is reminiscent of a power-law, and in the inset we plot the exponent of power-law fits. \emph{Bottom: } Distribution of the residence times of the water molecules present at the interface a the end of the simulation run. The black arrow points to the long-time bump in the distribution that is particularly visible at low hydration ($\lambda\leq 6$).
}
\label{fig:distributions}
\end{figure}

From the lists $\{p_{i,n}\}$ we can in principle determine the probability distribution of the residence times at the interface of all molecules, by simply counting the number of non-zero segments of given size (for each $i$). This should allow us to define an average survival time at the interface. Unfortunately this is problematic, as it is clear from the data sets shown in Fig.~\ref{fig:distributions} top, at the indicated values of $\lambda$. These distributions show significant power-law tails $\propto t^{-x}$ (solid lines in the main panel and points in the inset), with even $x<2$ at the lowest hydrations ($\lambda \leq 2$) which hinders a consistent definition of average values, and $2<x<3$ for $2<\lambda\le10$, which results in an undefined variance. This also implies that there is always a significant probability to observe life-times corresponding to the total time length of the simulation, making extremely hard adequate statistical sampling. This conclusion is verified from the data of Fig.~\ref{fig:distributions} (bottom), where we plot the same probability distributions as above, with the additional constraint that the considered non-zero segments must include $\tau_{\text{max}}$~\footnote{Note that the two probability distributions must coincide in the infinite-time limit.}. Clearly, at low hydration values the most likely residence time of water molecules present at the  interface at the end of the run is the duration of the run, as indicated by the arrow.

These observations are of particular interest, because long-tailed distributions of survival times are known to give rise to sub-diffusive random walks. This is the case, for instance, of the continuous-time random walk (CTRW), where particles are allowed to move after having been arrested for a waiting time, $t_w$, drawn from a power-law distribution. Evidences of CTRW have been observed in simulations of water confined in lipid bi-layers~\cite{Yamamoto2013,Yamamoto2014a}, among other cases. 

\begin{figure}[t]
\centering
\includegraphics{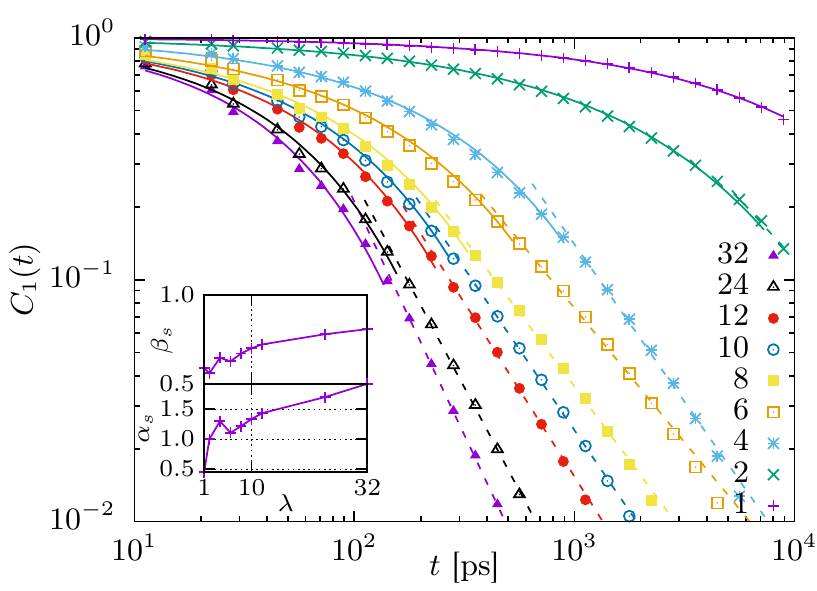}
\caption{
{\em Main panel:} Survival-time correlation functions of water at the interface, $C_1(t)$, computed via Eq.~(\ref{eq:corr-interface}), at the indicated values of $\lambda$. {\em Insets:} $\lambda$-dependence of the two fitting parameters $\beta_s$ and $\alpha_s$, obtained by representing the data with a modified exponential decay at short-times and a long-time power-law. The rationale for the fits shown as solid lines in the main panel is discussed in details in the main text.
}
\label{fig:time_interface}
\end{figure}

The difficulties described above can be partially circumvented by determining the time scale associated to the variation of well-designed correlation functions. We first consider the correlation function of the survival times of water molecules laying within the first hydration layer of the sulfonated groups as
\begin{equation}
C_1(t_n) = \frac{1}{1-n/N_s} \frac{\left\langle \sum_{s=0}^{N_s-1-n} (1-p_{i,n+s})\prod_{m=s}^{n+s-1} p_{i,m} \right \rangle_i}{\left\langle \sum_{s=0}^{N_s-1} p_{i,s} \right \rangle_i}.
\label{eq:corr-interface}
\end{equation}
Here, $\langle\rangle_i$ is the average over all water molecules $i=1,\ldots,N_{w}$, and $t_n=t_{n+s}-t_s$. Note that the product at the numerator is non-zero only if particle $i$ survives at the interface at all times $t_s\leq t < t_{n+s}$, subsequently detaching exactly at $t_{n+s}$. We plot our data in the main panel of Fig.~\ref{fig:time_interface} as symbols, at the indicated values of $\lambda$. 

Already by visual inspection it is clear that the typical lifetime of water molecules at the interface increases by lowering $\lambda$. Interestingly, we also find that two different mechanisms seem to concur at the relaxation on different time scales. Our data can be represented at all hydrations by a modified exponential, $C_1(t)\propto e^{-(t/\tau_s)^{\beta_s}}$ at short times, crossing-over to a power-law  behaviour, $C_1(t)\propto t^{-\alpha_s}$, at longer times. Note that the cross-over point depends on $\lambda$. The fitting curves are shown as solid and dashed lines in Fig.~\ref{fig:time_interface}. 

In our understanding, the faster process is associated to the exchange mechanism of water molecules which detach from the interface and transit in higher-order hydration layers, on a (well-defined) typical time scale $\tau_s(\lambda)$. Indeed, the functional form used is suggested by an analogy with the decay of a subset (population) of a given system, characterised by some particular feature. Similarly to the case of radioactive decay, for instance, if the probability to disintegrate (or to detach from the interface in our case) is uniform, then $C_1(t)\approx e^{-t/\tau_s}$. In contrast, if the probability to detach is not a constant, then $C_1(t)$ can relax both faster or slower than a simple exponential. The mildest deviation from the pure exponential is $C_1(t) \approx e^{-(t/\tau_s)^{\beta_s}}$, with the exponent $\beta_s$ quantifying the degree of the deviation: for $\beta_s<1$ (stretched) the decay is slower than exponential, faster for $\beta_s>1$ (compressed). In more extreme cases, scale free power-law behaviour have also been observed~\cite{Yamamoto2013,Yamamoto2014a}. 

Our analysis suggests that $\tau_s$ increases from $60$ to $22000$~ps as the hydration $\lambda$ decreases from $32$ to $1$ (see Fig.~\ref{fig:fits}), meaning that the exchange mechanism at the interface is activated by hydration. Also, we have found that $\beta_s$ varies from $0.6$ at $\lambda=1$ to $0.8$ at $\lambda=32$ (inset of Fig.~\ref{fig:time_interface}). These observations therefore seem to demonstrate that hydration controls both typical time-scale and statistical features of the postulated exchange mechanism. In addition, the self-similar behaviour observed at longer times clearly is the echo of the power-law tails of the static distribution functions of the $\{p_{i,n}\}$ of Fig.~\ref{fig:distributions}. It is natural to expect that this relaxation stems from trapping at the interface of water molecules on extremely extended time scales. This would also directly allow to identify interfaces as the main source of water sub-diffusion, rationalising this mechanism in terms of a CTRW approach. 
\begin{figure}[t]
\centering
\includegraphics[width=0.5\textwidth]{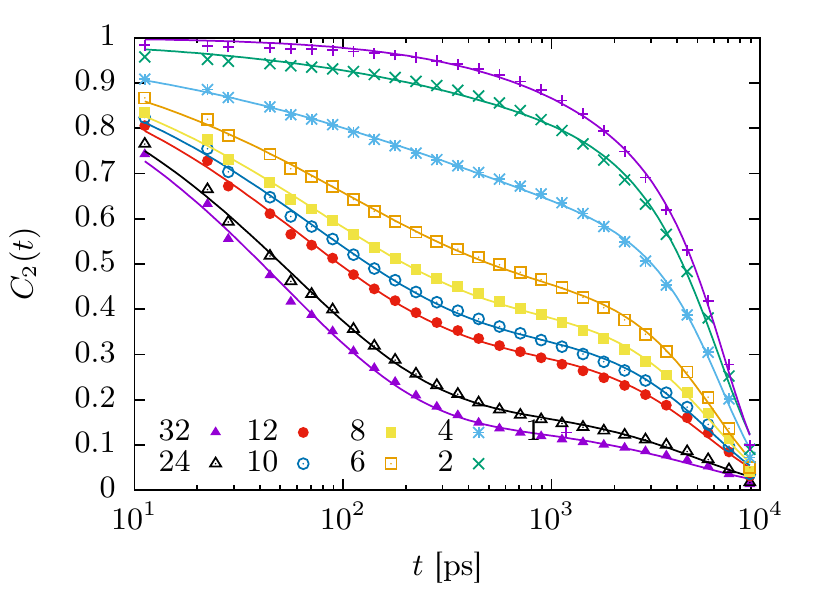}
\caption{
Time autocorrelation functions of $\{p_{i,n}\}$ (symbols), computed via Eq.~(\ref{eq:acf}), at the indicated values of $\lambda$. The continuous lines are the results of the fits to our data according to Eq.~(\ref{eq:fitacf}). A comprehensive discussion of these results is provided in the main text. 
}
\label{fig:acf-2exp}
\end{figure}

It is not possible, however, to associate a well-defined time scale to this dynamics in terms of Eq.~(\ref{eq:corr-interface}). Indeed, in order to highlight the dynamics most controlled by the interface, beyond the exchange process quantified above, one should lift the excessive constraint of water molecules {\em uninterruptedly} bounded to the interface during the time window $t_n$ implicit in that definition. We therefore define the time autocorrelation function of $\{p_{i,n}\}$,
\begin{equation}
C_2(t_n) = \frac{\left \langle \sum_{s=0}^{N_s-1-n} p_{i,s+n}p_{i,s} \right \rangle_i}{\left \langle \sum_{s=0}^{N_s-1} p_{i,s}^2 \right \rangle_i}.
\label{eq:acf}
\end{equation}
At variance with the case of Eq.~(\ref{eq:corr-interface}), with this definition water molecules that are at the interface at times $t_s$ and $t_{s+n}$ contribute to $C_2(t_n)$, notwithstanding their faith between $t_s$ and $t_{s+n}$, and at $t_{s+n+1}$. This function therefore still selects molecules that detach from the interface, but also those that return to the interface, at some later time after leaving. We expect, in general, that this dynamics takes place on time scales comparable to or longer than $\tau_s$. We plot our data as symbols in Fig.~\ref{fig:acf-2exp}, at the indicated values of $\lambda$.

The data show a two-relaxation pattern, that we describe as,
\begin{equation}
C_2(t) = c\, e^{-(t/\tau_i)^{\beta_i}} + (1-c)\, e^{-(t/\tau_\infty)^{\beta_\infty}}.
\label{eq:fitacf}
\end{equation}
We consider here two points. First, the initial decay of $C_2(t)$ takes place on a quite large range of time scales $\tau_i$, ranging from $\simeq 100$~ps at $\lambda=32$ to $\simeq 1$~ns at $\lambda=1$. This quantifies dynamics at the interface, as we will see below. Second, the functions exhibit an abrupt singular behaviour as $t$ approaches the total time window of our simulation, $\tau_{\text{max}}$, and this cutoff is more pronounced at lower hydrations. We indicate this limit time scale as $\tau_\infty$, which characterises molecules that are permanently attached to the interface on the time scale of our simulation. We postpone a comprehensive discussion of these data to the next Section.
\section{Discussion}
\label{sect:discussion}
\begin{figure}[t]
\centering
\includegraphics[width=0.5\textwidth]{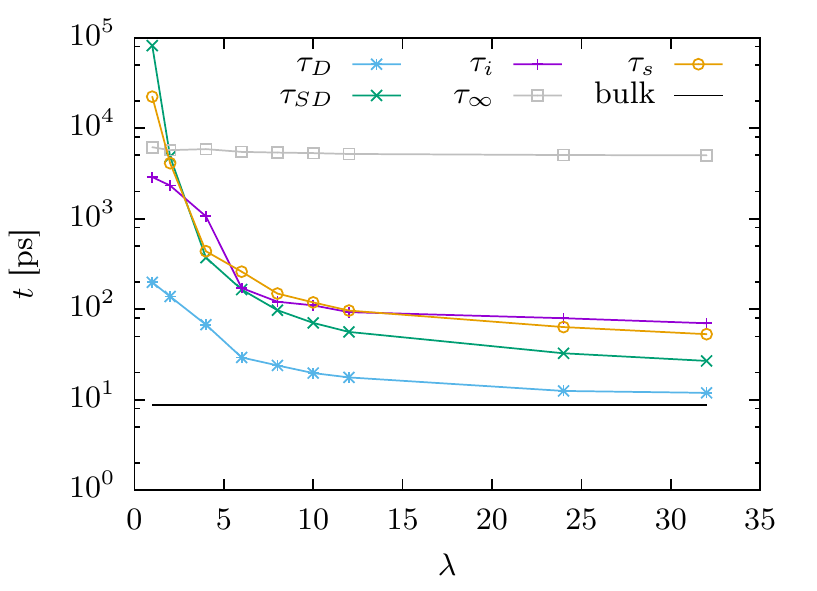}
\includegraphics[width=0.5\textwidth]{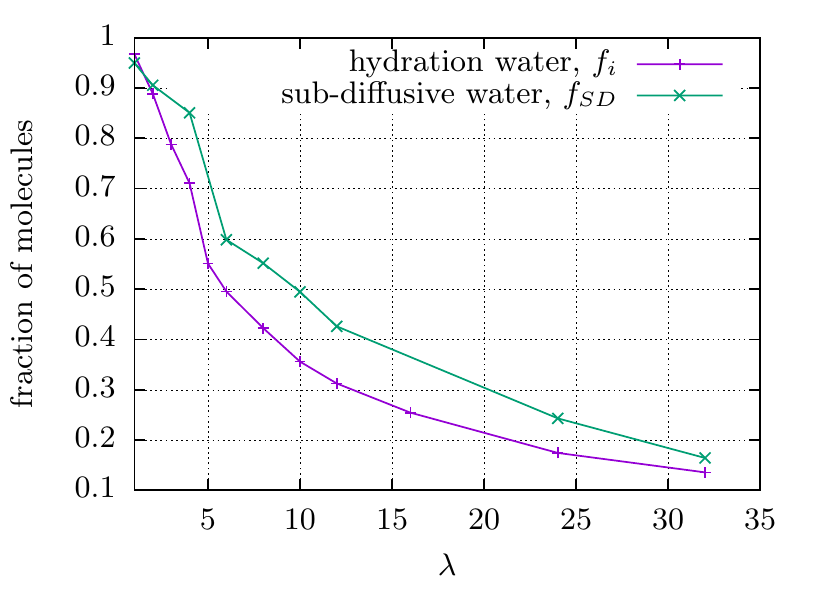}
\caption{
{\em Top:} 
$\lambda$-dependence of the relevant time-scales associated to the dynamics of the confined water molecules: $\tau_D$ and $\tau_SD$ are associated to the relaxation of the intermediate scattering function, $\tau_s$ characterises the survival-time correlation function, $C_1(t)$, $\tau_i$ and $\tau_\infty$ have been determined from the binary presence autocorrelation function, $C_2(t)$. A comprehensive discussion of these data is included in the main text. {\em Bottom:} Static fraction of water molecules residing at the interface, $f_i$, together with the dynamic fraction of sub-diffusive molecules, $f_{SD}$. The two data-sets coincides at low hydration. This point, together with the discrepancy at high-$\lambda$ are described in details in the main text.}
\label{fig:fits}
\end{figure}

We now describe the implications of the data reported in the previous Section, and provide a consistent unique picture of our observations. From the analysis above, we have been able to extract five time-scales relevant for the confined dynamics of the water molecules. In particular, $\tau_D$ and $\tau_{SD}$ are associated to the relaxation of the intermediate scattering function, $\tau_s$ characterises the survival-time correlation function $C_1(t)$, $\tau_i$ and $\tau_\infty$ have been determined from the binary presence autocorrelation function $C_2(t)$. We plot all data as a function of the hydration level $\lambda$ in the top panel of Fig.~\ref{fig:fits}. 

We see that the long-time component of the autocorrelation of the binary presence, $\tau_\infty$, is $\lambda$-independent, and of the order of the total simulation time. We associate this time-scale to the water molecules that stay in the interfacial layer for times that are compatible with the duration of the entire simulation run. As already anticipated, this is an artificial feature which originates from arbitrarily fixing a bound from above on the available observation time. As a consequence, all dynamics on time scales larger than $\tau_\infty$ cannot be probed in equilibrium in our simulation.

Interestingly, the time-scales $\tau_i$ of the short-time component of $C_2(t)$ and $\tau_s$ of the initial decay of $C_1(t)$ are, at variance, strongly reduced as the water content increases. We can associate these relaxations to an exchange mechanism between the interfacial region and the rest of the aqueous domains, which is activated by the increase of the hydration. This idea is strongly supported by a surprising observation: $\tau_i$ and $\tau_s$ are very similar, especially at low-$\lambda$, to the time scale $\tau_{SD}$ that we extracted from the $F_s(q^*,t)$, and associated with the sub-diffusive fraction of water molecules. These therefore seem to be those pertaining to the population with a dynamics of primary interface-dominated character.

An other observation additionally strengthen our picture. In Fig.~\ref{fig:fits} (bottom) we show two quantities: the {\em static} fraction over the total number of water molecules lying at the interface, $f_i$, which we obtain by simply counting the number of molecules comprised in the hydration layer; and the {\em dynamic} fraction of sub-diffusive particles, $f_{SD}$, obtained from Eq.~(\ref{eqn:fqt-fit}). The two data sets exhibit very similar variation on the whole hydration range, and the two curves actually coincide at the lowest hydration ranges, $\lambda\leq 2$. At higher hydration levels, $f_{SD}$ is close to, although always higher than, $f_i$. We infer from the similarities of the typical time scales and associated fractions (over the total number) that interfacial and sub-diffusive populations coincide, or at least that the sub-diffusive population is a superset of the interfacial one.
\begin{figure}[b]
\centering
\includegraphics[width=0.5\textwidth]{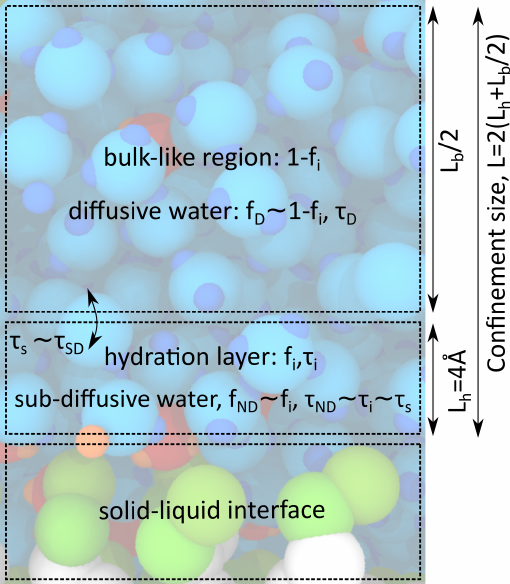}
\caption{
Schematic overview of the interplay between the structure of the surfactant phase and the transport properties of the confined water molecules. This sketch complements the detailed discussion contained in the main text.}
\label{fig:interface}
\end{figure}
\begin{figure*}[t]
\centering
\includegraphics[width=1.0\textwidth]{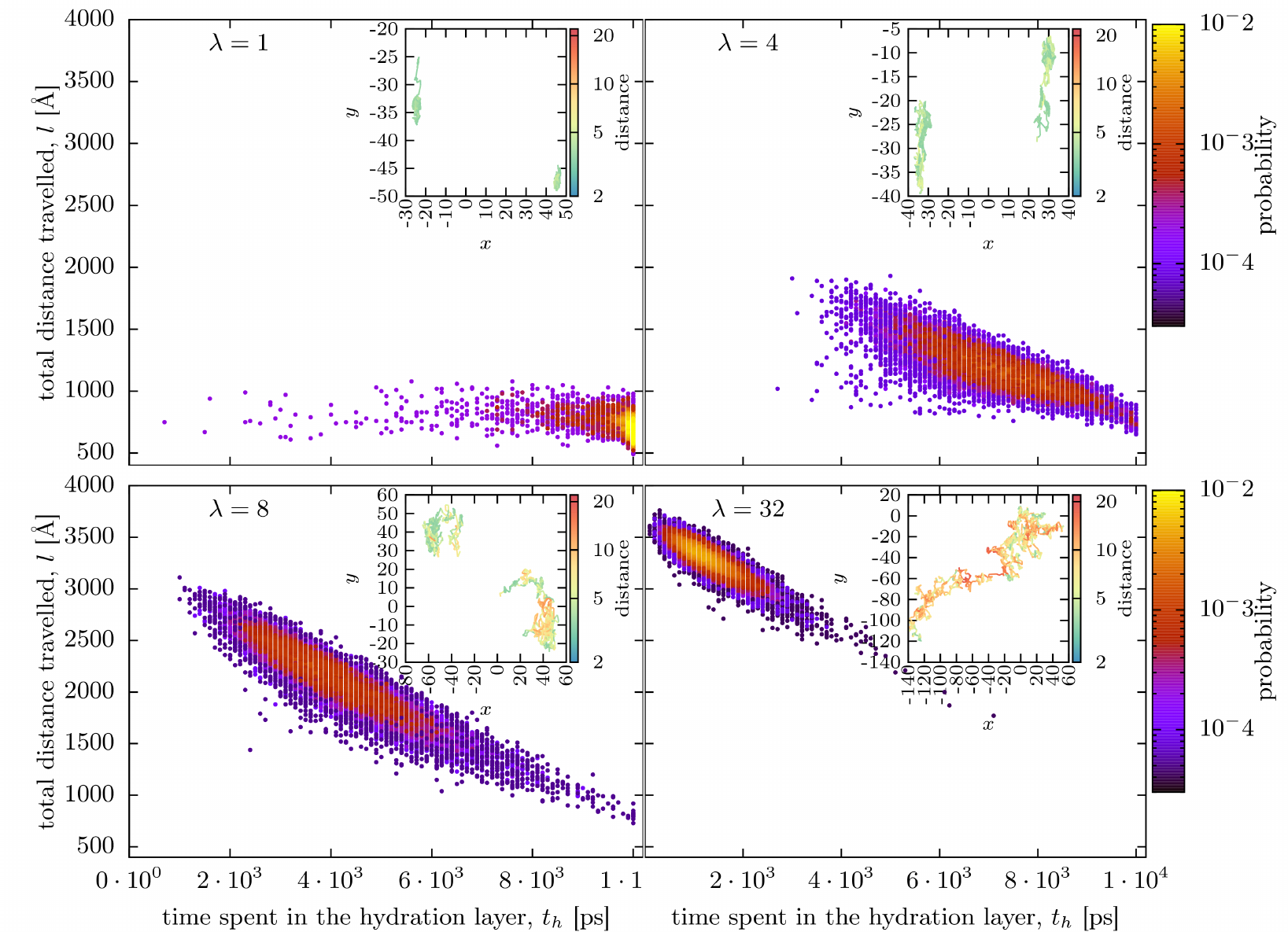}
\caption{
Joint probability distribution of the total displacement, $l$, of water molecules on the entire time scale of the run, $\tau_{\text{max}}$, and the total time, $t^h$, they have spent (even discontinuously) in the hydration layer of a surfactant head, on the same time window. The probability is represented as colour maps, for the indicated values of $\lambda$: yellow indicates high probability, indigo lower. In the insets we show the projection on the $xy$-plane of the trajectory followed by two molecules chosen at random at the corresponding $\lambda$, with the incremental distance travelled at each time indicated by the displayed colour code.
}
\label{fig:time_dist-Ow}
\end{figure*}

We propose a schematic overview of our discussion in Fig.~\ref{fig:interface}, where we partition into three main areas the surfactant phase environment, with both the confining matrix and the complementary space available for transport in the ionic channel, of typical size $\tilde{l}$: {\em i)} the solid-like interface (bottom), which comprises the surfactant hydrophilic heads; {\em ii)} the hydration layer (centre), that covers the interface and comprises all water molecules within $r_c=4$~\AA\, from the the nearest surfactant head; and {\em iii)} the bulk region (top) consisting of all remaining water molecules and extending over a typical length scale $l_b$, which depends on $\lambda$. Obviously, $\tilde{l}=2(l_h+l_b/2)$. 

The hydration layer is characterised by the fraction of molecules it comprises, $f_i$, together with the characteristic time of their interface-dominated dynamics, $\tau_i$. The fraction of water at the interface coincides at low-$\lambda$ (and has a very large overlap at high-$\lambda$) with the fraction of sub-diffusive water ($f_i\approx f_{SD}$), and one finds $\tau_i \approx \tau_{SD}$. The molecules at the boundary with the region extending beyond the first hydration shell and comprising the remaining fraction, $1-f_i\approx f_D$, participate to the exchange process with the molecules pertaining to the bulk-like reservoir. This takes place on a time scale $\tau_s$, which also correlates to the sub-diffusive time scale of the intermediate scattering function, with $\tau_s\approx \tau_{SD}$.

The sub-diffusive character of the mean-squared displacements (Fig.~\ref{fig:subdiffusion}) also finds a clear rationalisation in this picture. $\langle r^2(t) \rangle$ grasps an average dynamics, which results from a $\lambda$-dependent interplay of interface-dominated and bulk-like behaviour. The source of the observed sub-diffusion is trapping of water molecules at the interface, with a power-law distribution of trapping times, as demonstrated in Fig.~\ref{fig:time_interface}. This breaks the validity of the central limit theorem in a fashion reminiscent of a continuous time random walk, immediately providing a rigorous justification of the observed non-linearity of $\langle r^2(t) \rangle$ at long times. Also note that in general $f_i \leq f_{SD}$, and at high hydrations (above $\lambda=6$) $\tau_s$ and $\tau_i$ are slightly larger than $\tau_{SD}$. This indicates that water molecules exiting the hydration shell retain a certain degree of sub-diffusivity, with this anomalous character progressively disappearing when moving away from the interface.

To further clarify this point, we have quantified the correlation between the total displacement, $l_i=\sum_{n=1}^{N_s-1} |\mathbf{r_i}(t_n)-\mathbf{r_i}(t_n-1)|$ , of water molecule $i$ on the entire time scale of the run ($\tau_{\text{max}}\simeq 10$~ns) and the total time, $t^h_i$, it has spent (even discontinuously) in the hydration layer of a surfactant head, on the same time window. In Fig.~\ref{fig:time_dist-Ow} we show the joint probability distributions of these two quantities as colour maps, for the indicated values of $\lambda$; yellow indicates high probability, indigo lower. We observe a clear trend as the hydration increases.

At very low water content ($\lambda=1$), almost all water molecules remain in the hydration layer for most of the duration of the simulation run ($t_h\approx\tau_{\text{max}}$), and the total distance travelled does not strongly depend on the time spent at the interface, fluctuating around $l\approx 750$~\AA. In the inset we show the projection on the $xy$-plane of the trajectory followed by two molecules chosen at random, with the distance from the nearest O$_3$ bead indicated by the displayed colour code. The motion is clearly very constrained (probably bi-dimensional) in both cases, with a very limited portion of space explored. For $\lambda=4$ and $8$, however, the probability to observe very high values of $t_h$ progressively decreases, while the distribution of the $t_h$ becomes wider. This effect is accompanied by longer $l$ travelled at shorter $t_h$, with a degree of variation increasing with $\lambda$, as confirmed by the data in the insets. At $\lambda=32$, eventually, the probability accumulates in the high-$l$/low-$t_h$ region, indicating that practically all molecules transit between the interface and the bulk-like region (the trajectory of one molecule is shown in the inset, and clearly shows segments of high-speed travel far from the interface, and ``sticky'' regions that trap the molecule).

The conclusion arising from the above arguments is that the picture based on two well-identified dynamical population, that fully accounts for the behaviour in the low-$\lambda$ regime, must be modified at high water content, where we observe an increasing rate of the exchange mechanism and of the number of molecules participating to it. Indeed, it is not possible to account for the complex distributions of Fig.~\ref{fig:time_dist-Ow} in terms of an average $\tau_i$ only, and one should rather consider a $\lambda$-dependent set of populations, characterised by a distribution of $\tau_i$ ($\tau_{ND}$).
\section{Conclusions}
\label{sect:conclusion}
In this work we have followed numerically the dynamics of water molecules constrained in self-organised ionic surfactant phases. By varying the water content, we have considered an extended range of soft mobile confining environments, characterised by variable confining scales and topological properties of the generated hydrophilic/hydrophobic interfaces. These two features contribute in a highly non-trivial fashion to transport of the adsorbed fluid, that we have managed to characterise in terms of a space-dependent dynamical behaviour. 

More in details, we have demonstrated that both the observed sub-diffusive character of the average mean squared displacements, increasing by decreasing the length scale of the confinement, and the details of the relaxation of the intermediate scattering function, can be rationalised in terms of two dynamical populations. The first is predominantly interface-dominated, with a marked non-diffusive behaviour, the second has a bulk-like character. The average dynamical behaviour of the adsorbed fluid is primarily determined by the exchange mechanism between the two dynamical groups, whose precise features depend on the water content. This simple picture allows for a complete quantitative characterisation of dynamics at low-$\lambda$, while at high-$\lambda$ we have proposed that an entire $\lambda$-dependent set of populations must be considered to obtain the same accuracy. 

Overall, we have managed to provide, to the best of our knowledge, the first comprehensive picture of the dynamics of water molecules in soft confining environments, ranging from the case of extremely tight channels, to that where the interface is in contact with an infinite reservoir of bulk water. These two limits cover, we believe, most part of the typical situations encountered in hydrated natural and synthesised materials. Our model, as a consequence, could play for soft materials a critical paradigmatic role similar to that of carbon nanotubes for confinement in rigid ordered matrices.

\bibliography{references}
\end{document}